\documentclass[preprint]{aastex}

\usepackage{amsmath}

\shorttitle{Near-infrared line IDs}
\shortauthors{Friesen et al.}

\begin{document}

\title{Near-infrared line identification in type~Ia supernovae during the transitional phase}

\author{Brian~Friesen\altaffilmark{1}, E.~Baron\altaffilmark{1,2}, John~P.~Wisniewski\altaffilmark{1}, Jerod~T.~Parrent\altaffilmark{3}, R.~C.~Thomas\altaffilmark{4}, Timothy~R.~Miller\altaffilmark{1}, and G.~H.~Marion\altaffilmark{5}}

\altaffiltext{1}{Homer L. Dodge Department of Physics and Astronomy, 440 W. Brooks St., Room 100, Norman, OK 73019, USA}
\altaffiltext{2}{Hamburger Sternwarte, Gojenbergsweg 112, 21029 Hamburg, Germany}
\altaffiltext{3}{Harvard-Smithsonian Center for Astrophysics, 60 Garden Street, Cambridge, MA 02138, USA}
\altaffiltext{4}{Computational Cosmology Center, Computational Research Division, Lawrence Berkeley National Laboratory, 1 Cyclotron Road MS 50B-4206, Berkeley, CA 94720, USA}
\altaffiltext{5}{University of Texas at Austin, 1 University Station C1400, Austin, TX, 78712-0259, USA}

\begin{abstract}
    We present near-infrared synthetic spectra of a delayed-detonation hydrodynamical model and compare them to observed spectra of four normal type~Ia supernovae ranging from day +56.5 to day +85.
    This is the epoch during which supernovae are believed to be undergoing the transition from the photospheric phase, where spectra are characterized by line scattering above an optically thick photosphere, to the nebular phase, where spectra consist of optically thin emission from forbidden lines.
    We find that most spectral features in the near-infrared can be accounted for by permitted lines of \ion{Fe}{2} and \ion{Co}{2}.
    In addition, we find that [\ion{Ni}{2}] fits the emission feature near 1.98~$\mu$m, suggesting that a substantial mass of $^{58}$Ni exists near the center of the ejecta in these objects, arising from nuclear burning at high density.
    A tentative identification of \ion{Mn}{2} at 1.15~$\mu$m may support this conclusion as well.
\end{abstract}


\section{Introduction}
\label{sec:intro}

Type~Ia supernovae (SNe~Ia) radiate primarily at optical wavelengths.
However, optical and especially ultraviolet (UV) radiation is sensitive to attenuation by dust along the line of sight to the observer \citep{ccm89}.
Furthermore, these bands are extremely crowded regions of the SN spectral energy distribution (SED): generally over 100 lines lie within one Doppler width of one another at any given optical wavelength, with the number approaching 1000 in the UV \citep{baron96}.
The near-infrared (NIR) bands of a SN~Ia SED behave very differently than the UV or optical, having adjacent regions of wildly varying line opacity \citep{wheeler98}.
This phenomenon allows one to probe the ejecta at a wide range of velocities, providing a unique window to the radiation physics of SNe~Ia, which in turn can constrain properties of explosion and progenitor scenarios.
Spectroscopic studies of NIR spectra of SNe~Ia at both early epochs \citep[e.g.,][]{meikle96,wheeler98,hsiao13} and very late epochs \citep{spyromilio92,bowers97,hoeflich04,mazzali11} have provided wide-ranging constraints on the distribution of unburned and partially burned nuclear white dwarf (WD) material, as well as the explosion energetics and $^{56}$Ni mass.
Several other studies have focused on a time-series of NIR spectra which span from pre- to post-maximum light, e.g., \citet{hamuy02,hoeflich02,marion03,marion09,gall12}; these studies have generally reached similar conclusions about post-maximum NIR spectra, that they are formed primarily from iron-peak elements such as Fe and Co.

In this work we expand upon these studies by identifying the features in the NIR spectra of four normal SNe~Ia during the ``transitional'' phase, a few months post-maximum light, during which the ejecta are neither photospheric nor entirely nebular.
One of these objects is the recent and nearby SN~2014J, discovered by \citet{fossey14} in the nearby galaxy M82 approximately 1~week after explosion.
Early and extensive follow-up campaigns revealed that SN~2014J is a normal SN~Ia, but the environment and/or the line of sight to the explosion are highly reddened by dust \citep{goobar14,zheng14,marion14}.
The other three SNe examined in this work are SNe~2001fe \citep{01fe_fc}, 2002bo \citep{benetti04}, and 2003du \citep{gerardy04,hoeflich04,anupama05,stanishev07,tanaka11}.
The spectroscopic and photometric properties of these four SNe are summarized in Table~\ref{objsum}; all four are relatively normal SNe, with the exception of some high-velocity features in SNe~2002bo and 2014J.
Because several objects comprise our ``time-series'' of NIR spectra, we do not seek in this work to characterize the time-evolution of any particular spectroscopic feature, as it is difficult to separate the intrinsic variability of the ejecta of each SN from age-dependent effects in a given spectrum.
Rather, our purpose is to identify common features in each spectrum, and to infer properties inherent to our entire sample which may provide useful diagnostic tools in the study of larger samples or of a high-cadence time-series of NIR spectra of a single object which includes these epochs.
In Section \ref{sec:obs} we describe our observations of SN~2014J; in Section \ref{sec:num_mod} we present the numerical methods involved in calculating synthetic spectra; in Section \ref{sec:discussion} we compare our results to the NIR observations; and finally in Section \ref{sec:conclusions} we summarize our work.

\section{Observations}
\label{sec:obs}

We obtained two observations of SN 2014J with TripleSpec, a cross-dispersed infrared spectrograph mounted on the 3.5~meter telescope at Apache Point Observatory on 2014 April 8 and 2014 April 11.
At this epoch SN~2014J had NIR apparent magnitudes of approximately $J = 12.0$, $H = 10.6$, and $K = 10.8$ \citep{foley14}.
These data are the first of a pilot project to generate a wide-ranging data set of NIR spectra and photometry of nearby ($z \lesssim 0.008$) SNe~Ia at post-maximum and late-time epochs using TripleSpec.
The observations were made with the 1$\farcs$1 x 43$\farcs$0 slit, yielding resolution $R \sim 3500$ data from 0.95~$\mu$m~--~2.46~$\mu$m, and were obtained using a standard ABBA nod sequence.
Observations of a nearby A0V star were obtained immediately after our  observations of SN 2014J to facilitate accurate telluric subtraction.
Table \ref{obssum} summarizes the details of our observations, including the individual exposure times and total on-source integration times for SN 2014J and our A0V calibrators.

These data were flat-fielded, wavelength calibrated, and extracted using Triplespectool, a modified version of Spextool \citep{cushing04}.
The data were corrected for telluric features using observations of nearby A0V stars at similar airmass to SN 2014J following the techniques  outlined in \citet{vacca03}.
We also performed absolute flux calibration using the A0V telluric stars, following \citet{vacca03}: when combining the individual SN and stellar spectra, we scaled each to the mean computed in high signal-to-noise regions of the spectra.
The relative fluxes presented here are unaffected by these processes.
Our two SN~2014J observations were obtained at significantly different hour angles and airmasses.
As optimal removal of telluric features is achieved by minimizing the angular distance and airmass difference between science and A0V calibrator \citep{vacca03}, we utilized different A0V calibrators for our two epochs of observations (see Table~\ref{obssum}).
The spectra have been deredshifted by $v = 203$~km~s$^{-1}$ using M82 data from NED\footnote{The NASA/IPAC Extragalactic Database (NED) is operated by the Jet Propulsion Laboratory, California Institute of Technology, under contract with the National Aeronautics and Space Administration.} and dereddened using $R_V=1.7$ and $E(B-V)=1.2$~mag for M82 and $R_V=3.1$ and $E(B-V)=0.14$~mag for the Milky Way, following the estimates of \citet{goobar14}, although the reddening effects of dust from both galaxies are negligible in the NIR.

\begin{figure}
    \epsscale{1.0}
    \plotone{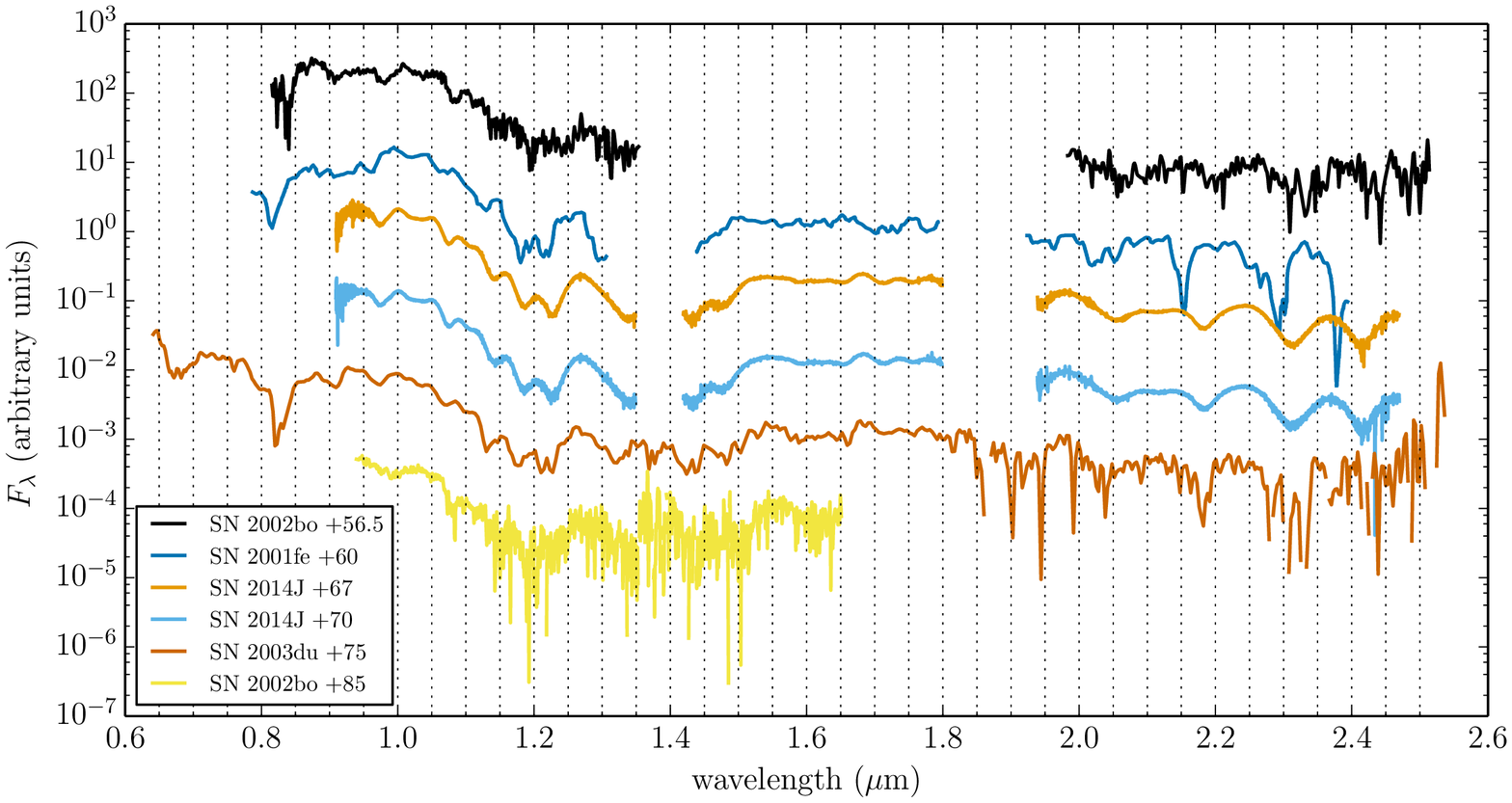}
    \caption{NIR spectra of SN~2014J compared with SNe~2001fe, 2002bo, and 2003du at similar epochs.
        All of the spectra have been deredshifted and dereddened.
        They have also been multiplied by arbitrary constants to facilitate comparison.}
    \label{fig:01fe_03du_14J}
\end{figure}

For comparison we plot alongside our SN~2014J spectra four transitional-phase NIR spectra: SN~2001fe at day +60, SN~2003du at day +75, and SN~2002bo at days +56.5 and +85 \citep{benetti04,marion09}.
The spectra of SN~2002bo were obtained from WISeREP \citep{wiserep}.
All six spectra are shown in Figure~\ref{fig:01fe_03du_14J}, with the flux of each multiplied by an arbitrary factor to facilitate comparison of spectral features.

\section{Numerical models}
\label{sec:num_mod}

\begin{figure}
    \epsscale{1.0}
    \plotone{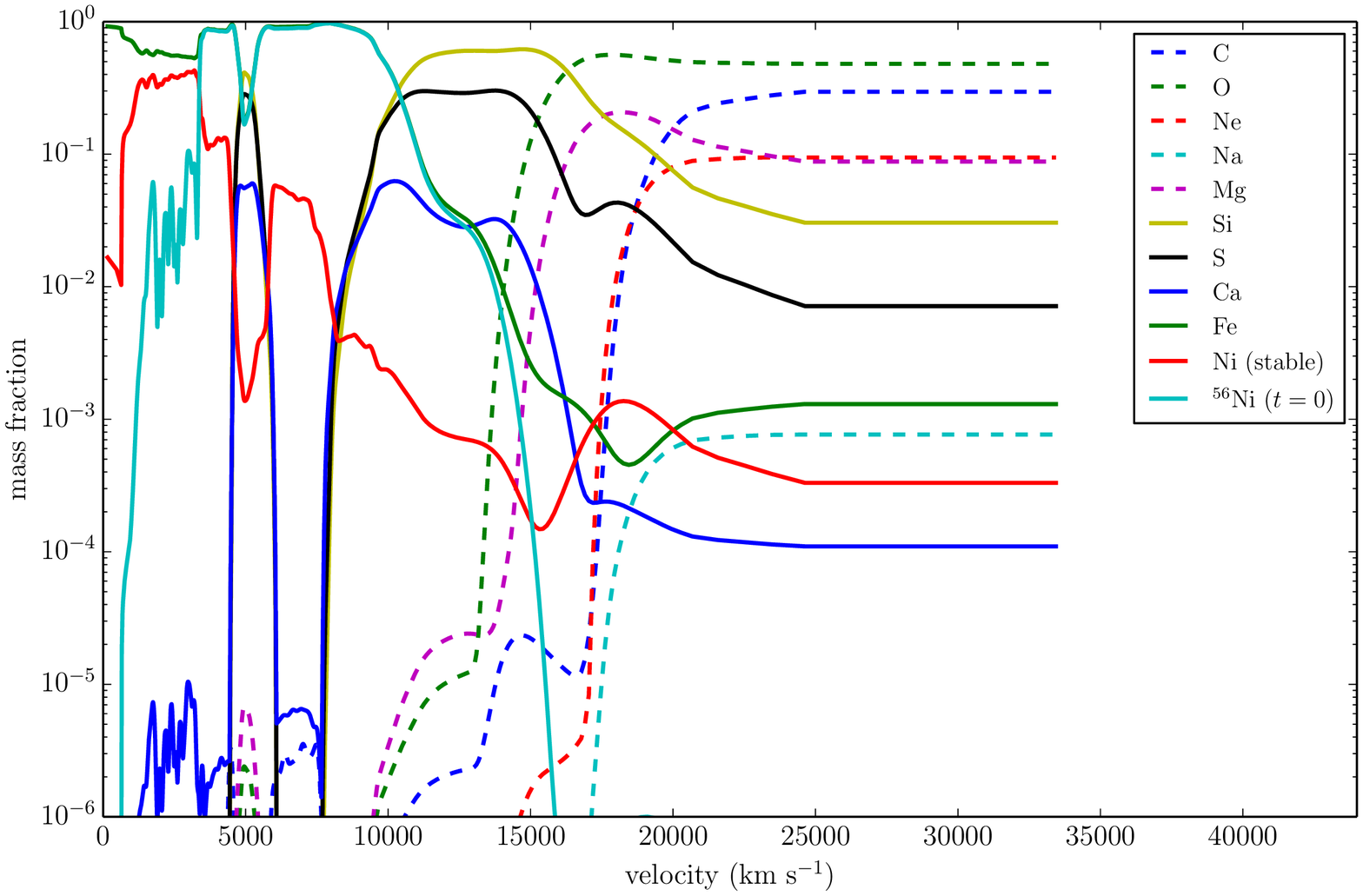}
    \caption{Abundances of delayed detonation explosion model used in this work.
        The $^{56}$Ni abundance is plotted at $t = 0$; all others are plotted at $t \rightarrow \infty$.
    The details of the explosion model are reviewed in Section \ref{sec:num_mod} and are described in more detail in \citet{dominguez01}.}
    \label{fig:pah_std}
\end{figure}

To generate model spectra we used the radiative transfer code \texttt{PHOENIX/1D} \citep[e.g.,][]{nugent95,hb99,lentz00,baron06,baron08}, which solves the spherically symmetric, special relativistic radiative transfer equation, correct to all orders in $\mathcal{O}(v/c)$, along with the non-local thermodynamic equilibrium (NLTE) equations of statistical equilibrium.
It also accounts for non-thermal processes such as $\gamma$-ray and positron deposition, and includes a complete treatment of line blanketing.
Our models contain approximately 100,000 wavelength points.
We calculated synthetic spectra of a spherically symmetric delayed-detonation explosion model described in \citet{dominguez01}, which reproduces the light curves and spectra for core-normal supernovae \citep{hoeflich95,hoeflich02,hoeflich06}.
In this realization, the carbon/oxygen WD is from the core of an evolved 5~$M_\odot$ main-sequence star.
Through accretion, this core approaches the Chandrasekhar mass, and an explosion begins spontaneously when the core has a central density of $2.0 \times 10^9$~g~cm$^{-3}$ and a total mass of 1.36~$M_\odot$.
The transition from deflagration to detonation is triggered artificially at a density of $2.3 \times 10^7$~g~cm$^{-3}$.
The model abundances are shown in Figure~\ref{fig:pah_std}, with $^{56}$Ni plotted at $t = 0$ and all other species plotted at $t \rightarrow \infty$ such that, e.g., the mass of Fe includes the stable Fe produced at $t = 0$ as well as all of the decayed $^{56}$Ni.
The explosion produces 0.64~$M_\odot$ of $^{56}$Ni and 0.09~$M_\odot$ of stable $^{58}$Ni.

We generated a unique synthetic spectrum to correspond to each of the six observed spectra, assuming each SN had a rise time of 15~d.
For example, the synthetic spectrum corresponding to the day +67 spectrum of SN~2014J had a homologous expansion time of 82~d.
While it is unlikely that all six SNe indeed had rise times of that period, the variation in rise times (likely a few days) should not affect the spectra significantly, given how slowly they evolve at these epochs \citep{marion09}.
Furthermore, our purpose here is to identify spectral features, not to quantify the time-evolution of a given SN (a difficult task given the dearth of available post-maximum NIR spectra).

Only two significant differences exist between the numerical methods used in previous \texttt{PHOENIX} calculations and the ones presented here.
In earlier works, we enforced radiative equilibrium via the modified Uns\"old-Lucy algorithm \citep{hb99}, where the total bolometric luminosity in the observer's frame is specified at the outer boundary and the $\mathcal{O}(v/c)$ moment equations are integrated to obtain the values of the radiation energy density $E$ and flux $F$ in the comoving frame.
Here, we have instead chosen to solve directly the time-independent $\mathcal{O}(v/c)$ radiation energy equation \citep{mm84},
\begin{equation}
    \label{eq:energy_eqn}
    \frac{1}{r^2} \frac{\partial}{\partial r} (r^2 F) + \frac{v}{r} (3 E - P) + \frac{\partial v}{\partial r} (E + P) = \int_0^\infty d\lambda (4\pi \eta_\lambda - c \chi_\lambda E_\lambda) - \frac{\dot{S}}{4\pi},
\end{equation}
where $E$, $F$, and $P$ are the radiation energy density, flux, and pressure, respectively, integrated over all wavelenths; $\eta_\lambda$ is the emissivity, $\chi_\lambda$ is the total opacity (absorption and scattering), and $\dot{S}$ is the rate of deposition of non-thermal energy due to radiative decay of $^{56}$Ni and $^{56}$Co, in the form of $\gamma$-rays and positrons.
By solving Equation~\ref{eq:energy_eqn} directly we do not specify a target luminosity at all.
In this way we enforce energy conservation (heating~=~cooling) and the luminosity is given by the total emission from the entire ejecta which is now powered by radioactive energy deposition (contained in $\dot{S}$ in Equation~\ref{eq:energy_eqn}).
The second difference is that here we have treated all line opacity in full NLTE, rather than parameterizing the opacity for weak lines as discussed in \citet{hauschildt95}.

Every term in Equation~\ref{eq:energy_eqn} except for $\dot{S}$ is a function of the gas temperature $T$.
Furthermore, the temperature-dependence of each of these terms is highly non-linear; therefore this equation requires iteration to solve.
To do so, we first choose two successive temperatures guesses for each point in the ejecta.\footnote{Because the temperature-dependent quantities in Equation~\ref{eq:energy_eqn} are obtained iteratively, it is difficult to obtain analytic temperature derivatives which are required for Newton-Raphson iteration.
        We therefore are relegated to using the secant method, which follows the same procedure as the Newton-Raphson method except with numerical derivatives.
        The disadvantage of the secant method is that it requires two initial conditions instead of just one.
        However, as long as both initial guesses are near the correct solution, the converged solution is independent of their values.}
        We next iterate the non-local thermodynamic equilibrium (NLTE) rate equations, approximate $\Lambda$-iteration, and the equation of state until all three are globally converged for the given temperature.
Then we apply the secant method to Equation~\ref{eq:energy_eqn} to obtain a new temperature guess.
Finally, we repeat the latter two steps until Equation~\ref{eq:energy_eqn} is satisfied\footnote{We consider a solution ``converged'' if $\Delta T/T $ is below 1\%.} at all points in the ejecta.

\section{Discussion}
\label{sec:discussion}

The observed NIR spectra shown in Figure~\ref{fig:01fe_03du_14J} are fairly similar.
However there are a few interesting differences, including the lines from 1.9~$\mu$m~--~2.5~$\mu$m in SN~2001fe and SN~2003du which are narrow and slightly blueshifted compared to SNe~2002bo and 2014J.
In addition, SN~2003du also may exhibit \emph{two} emission features between 2.3~$\mu$m -- 2.4~$\mu$m (although it is not easy to distinguish this possibility from the noise), while the other three SNe show only a single feature.
If it is indeed two distinct features, it is possible that they are present in all three SNe, and that only in SN~2003du are the velocities low enough for them to appear unblended.
If this were the case, one might attribute the lower velocity to an age effect, since SN~2003du's $K$-band spectrum is the oldest of those shown in Figure~\ref{fig:01fe_03du_14J}.
However, we believe this explanation is unlikely, given how slowly NIR spectra appear to change in the catalog of \citet[][cf. their Figure~9]{marion09}.
An alternative explanation is that the ejecta in the explosion of SN~2003du had a different $^{56}$Ni distribution than in SN~2001fe or SN~2014J, or that the velocity gradient of the line-emitting region is narrower in this object than in the others.

\begin{figure}
    \epsscale{1.0}
    \plotone{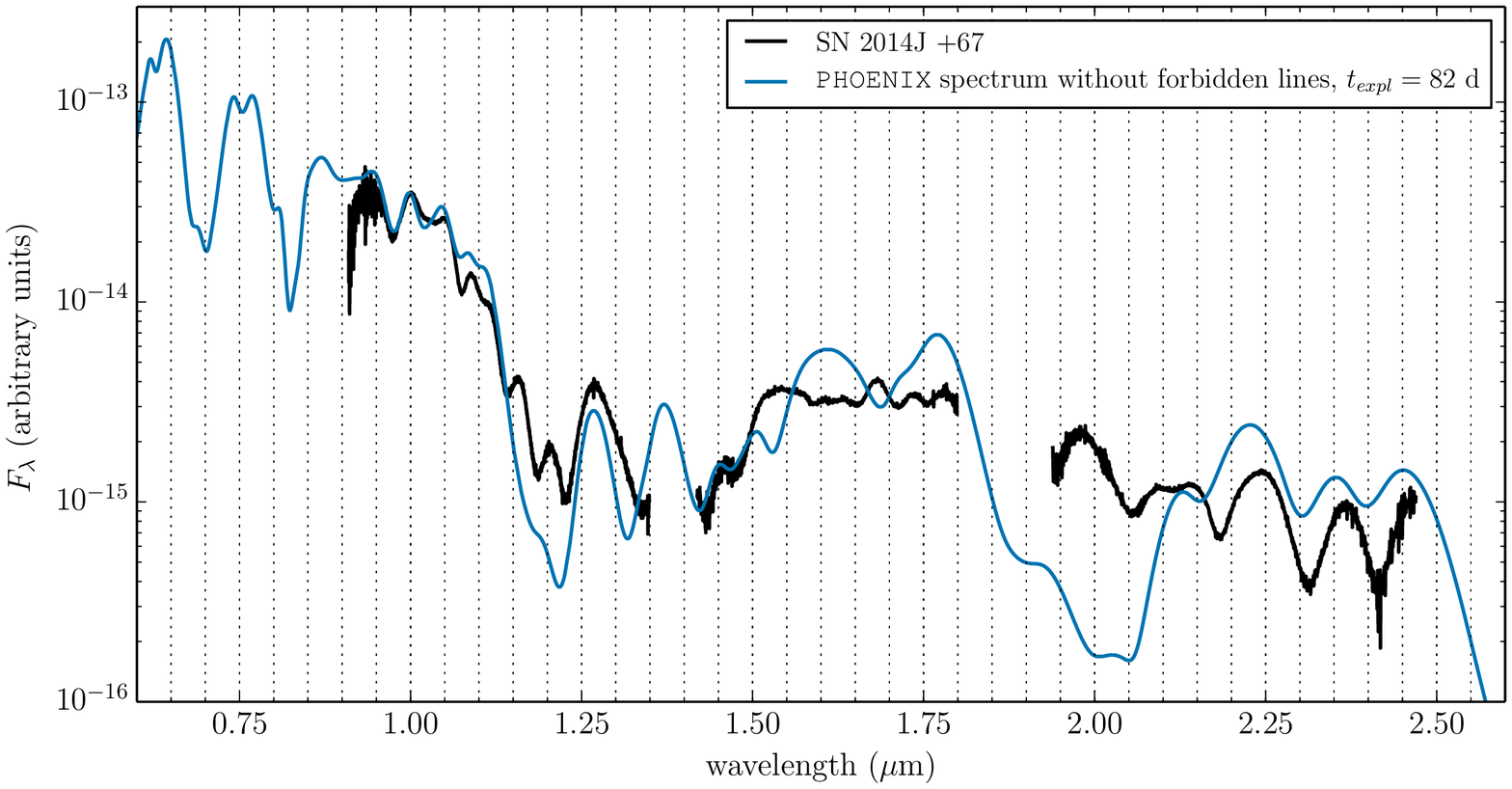}
    \caption{\texttt{PHOENIX} spectrum with only permitted lines, with a homologous expansion time of 82~d, compared to SN~2014J at day +67.
    The synthetic spectrum has been scaled to match the observed flux at 1.00~$\mu$m.}
    \label{fig:phx_no_forb_vs_14J}
\end{figure}

Figure~\ref{fig:phx_no_forb_vs_14J} shows the \texttt{PHOENIX} synthetic spectrum compared to SN~2014J at day +67, with the model spectrum scaled such that the predicted flux matches the observation at 1.00~$\mu$m.
(We apply the same scaling procedure in all comparisons of models and observations.)
The calculation used to generate this spectrum included no forbidden lines.
Most features in the synthetic spectrum match those in the observed spectrum quite well.
The primary shortcoming in the synthetic spectrum is the emission feature at 1.98~$\mu$m in SN~2014J which is not reproduced in the model.

In order to determine which atomic species and lines are responsible for which spectral features, we calculate ``single-ion spectra'' using the prescription of \citet{bongard08}.
The resulting spectra for the most influential ions are shown in Figure~\ref{fig:phx_single_ion_no_forb_vs_14J}.
These single-ion spectra indicate that nearly every feature in the NIR spectrum of SN~2014J can be accounted for with permitted lines of \ion{Fe}{2} and \ion{Co}{2}.
In particular, \ion{Fe}{2} dominates the NIR spectrum blueward of 1.50~$\mu$m, and \ion{Co}{2} dominates the spectrum redward of that point.
This result agrees with the conclusions of \citet{branch08}, who focused on the optical wavelengths of normal SN~Ia SEDs during the transitional phase, and \citet{gall12}, who showed that fluorescence of iron-peak and intermediate-mass elements produces much of the post-maximum NIR spectrum in the normal SN~2005cf.

Anticipating the possibility that forbidden lines play a role in NIR spectrum formation at these epochs, we have repeated the calculations of this model, including the forbidden lines in \citet{lundqvist96} and the NIST Atomic Spectral Database \citep{nist} for all species with $Z = 6$ through $Z = 28$ (carbon through nickel).
Comparing the permitted-line-only and the permitted-and-forbidden line spectra will then reveal the role that forbidden lines play during these epochs and in these NIR bands.
Because changing the atomic data changes the problem that is actually solved \citep[forbidden emission acts as an extra coolant which affects the temperature of the ejecta;][]{dessart14}, we re-converged the entire model after adding these extra lines.
The resulting spectrum, along with the corresponding single-ion spectra, is shown in Figure~\ref{fig:phx_single_ion_forb_vs_14J}.
We also show the corresponding temperature structure $T(v)$ for both models in Figure~\ref{fig:temps} in order to illustrate the cooling effects of forbidden lines.
We calculated a synthetic spectrum corresponding to each of the six observed spectra, which together are plotted in Figure~\ref{fig:phx_forb_vs_all_obs}.
The addition of forbidden lines does lower the temperature, as discussed in \citet{dessart14}, but the effect is fairly small, even at this epoch.

The few features which Fe and Co alone do not produce, merit additional scrutiny.
These include the emissions at 1.15~$\mu$m, 1.20~$\mu$m, and 1.98~$\mu$m.
We discuss identifications of each of these below.

\subsection{The 1.15~$\mu$m emission}

Figure~\ref{fig:phx_single_ion_no_forb_vs_14J} suggests that the emission at 1.15~$\mu$m could be due to a combination of \ion{Fe}{2} and \ion{Ca}{2}, \ion{Mn}{2}, and \ion{Ni}{2}.
Our model atoms, comprised mostly from the sources listed in \citet{hauschildt95} and \citet{short99}, predict 52 \ion{Mn}{2} lines, 56 \ion{Ni}{2} lines, 7 \ion{Ca}{2} lines, and 30 \ion{Fe}{2} lines, all within the range 1.14~$\mu$m -- 1.16~$\mu$m; more detailed level population analysis would be required to identify which of these is forming the emission seen in the single-ion spectra.
That the composite synthetic spectrum does not reproduce these features of the intermediate mass elements very well is possibly due to deficiencies in the hydrodynamical model or in the gas temperature calculation.

Adding forbidden lines to the calculation complicates the identification of this emission.
Figure~\ref{fig:phx_single_ion_forb_vs_14J} shows that the strongest components at that wavelength after \ion{Fe}{2} are now \ion{Co}{2} and \ion{Ni}{2}, followed by \ion{Mn}{2} and \ion{Ca}{2}.
Because the composite \texttt{PHOENIX} spectrum does not fit that feature very well, it is difficult to say which of these is responsible.

\subsection{The 1.20~$\mu$m emission}

In the permitted-line only spectrum (Figure~\ref{fig:phx_single_ion_no_forb_vs_14J}), the emission near 1.20~$\mu$m is likely due to a blend of \ion{Fe}{2} lines (there are 38 within in the range 1.19~$\mu$m -- 1.21~$\mu$m), \ion{Ca}{2}, (there are 6 within the same range), and \ion{Ni}{2} (45 lines).
As with the 1.15~$\mu$m feature, inclusion of forbidden lines complicates identification of this emission.
A number of \ion{Ni}{2} lines strengthen at nearby wavelengths, but \ion{Ca}{2} remains the strongest component after \ion{Fe}{2}.

\subsection{The 1.98~$\mu$m emission}

\begin{figure}
    \epsscale{1.0}
    \plotone{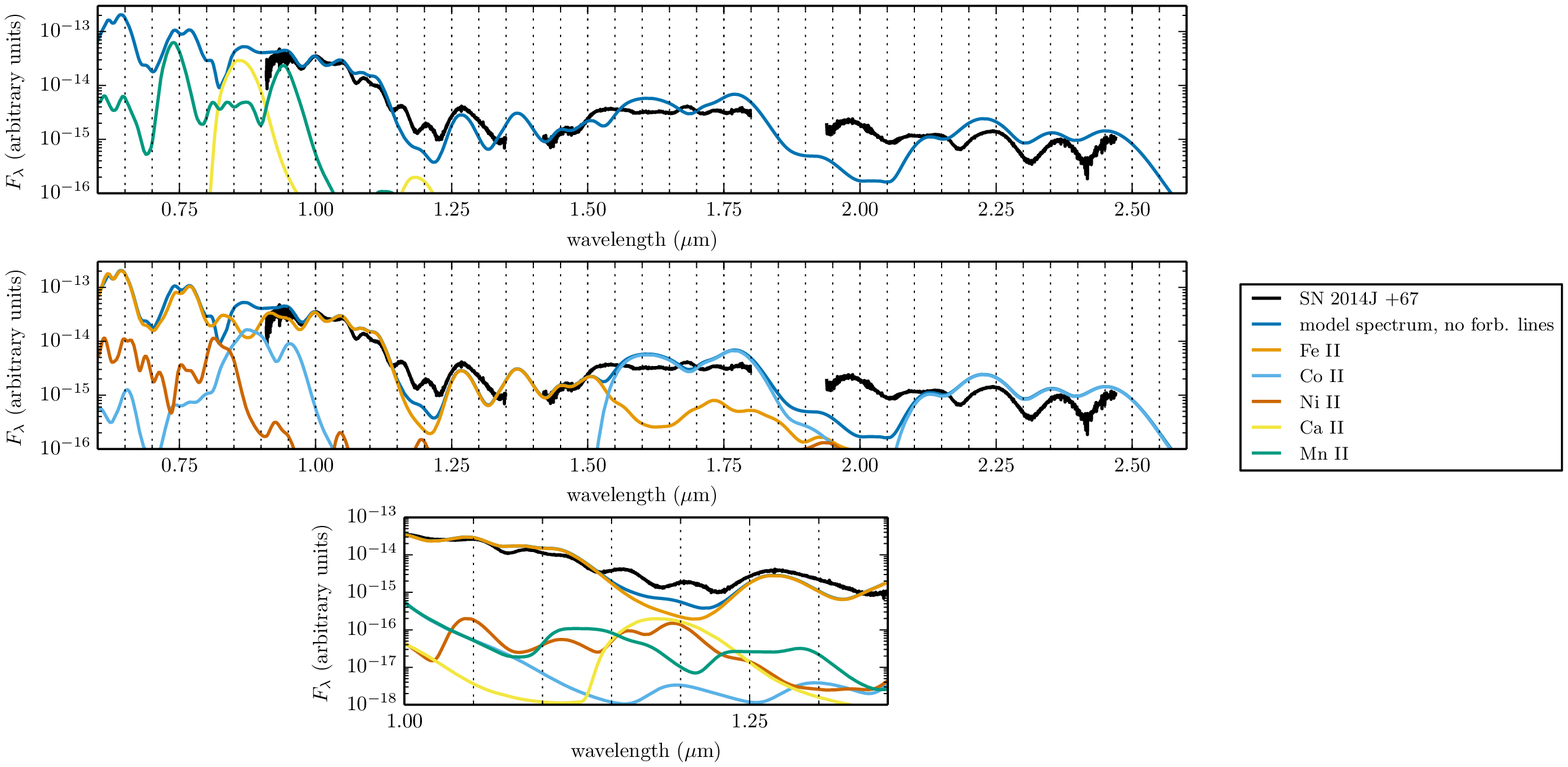}
    \caption{Single-ion spectra for the most influential ions in the \texttt{PHOENIX} calculation which did not include forbidden line data, compared to the composite synthetic spectrum as well as to SN~2014J at day +67.
        The top panel shows intermediate-mass elements; the middle panel shows iron-peak elements; the bottom panel shows an enlarged version of the emission features at 1.15~$\mu$m and 1.2~$\mu$m, which likely arise due to a combination of several ions.
        Most features blueward of 1.50~$\mu$m are fit with permitted \ion{Fe}{2}, while most features redward of that point are fit with permitted \ion{Co}{2}.
        The emission feature at 1.15~$\mu$m may be due to \ion{Mn}{2}, while that a 1.20~$\mu$m may be due to \ion{Ca}{2}.
        Permitted lines alone appear unable to fit the emission at 1.98~$\mu$m.}

    \label{fig:phx_single_ion_no_forb_vs_14J}
\end{figure}

\begin{figure}
    \epsscale{1.0}
    \plotone{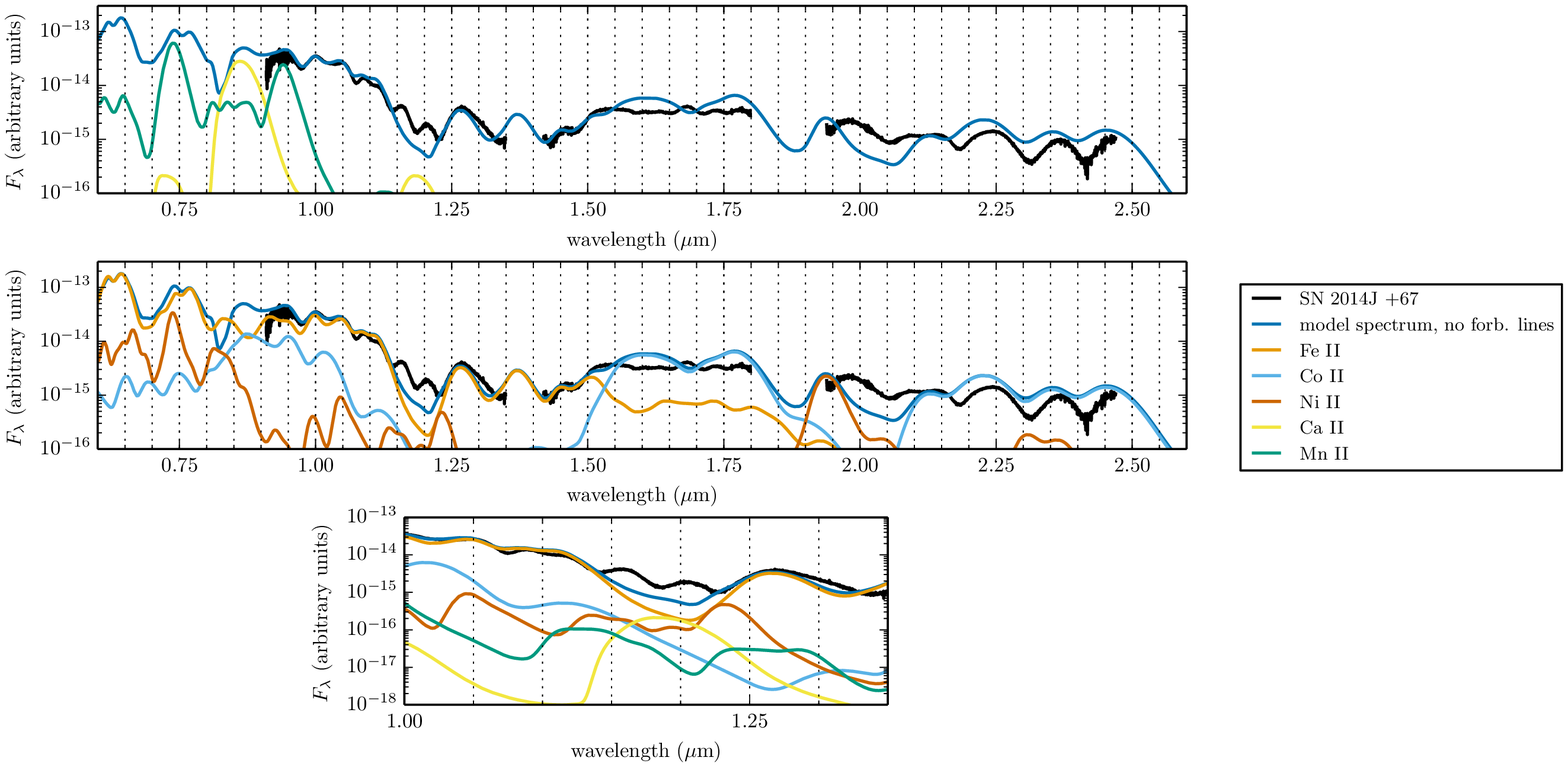}
    \caption{Single-ion spectra for the most influential ions in the \texttt{PHOENIX} calculation which included forbidden line data, compared to the composite synthetic spectrum as well as to SN~2014J at day +67.
        The top panel shows intermediate-mass elements; the middle panel shows iron-peak elements; the bottom panel shows an enlarged version of the emission features at 1.15~$\mu$m and 1.2~$\mu$m, which likely arise due to a combination of several ions.
        As in Figure~\ref{fig:phx_single_ion_no_forb_vs_14J}, most features are best fit by permitted \ion{Fe}{2} and \ion{Co}{2}.
        \ion{Mn}{2} may still be responsible for the feature at 1.15~$\mu$m, although now it appears that \ion{Co}{2} and \ion{Ni}{2} contribute as well.
        \ion{Ca}{2} remains the most likely candidate for the emission at 1.20~$\mu$m.
        The emission at 1.98~$\mu$m is due to [\ion{Ni}{2}], from stable $^{58}$Ni.}
    \label{fig:phx_single_ion_forb_vs_14J}
\end{figure}

\begin{figure}
    \epsscale{1.0}
    \plotone{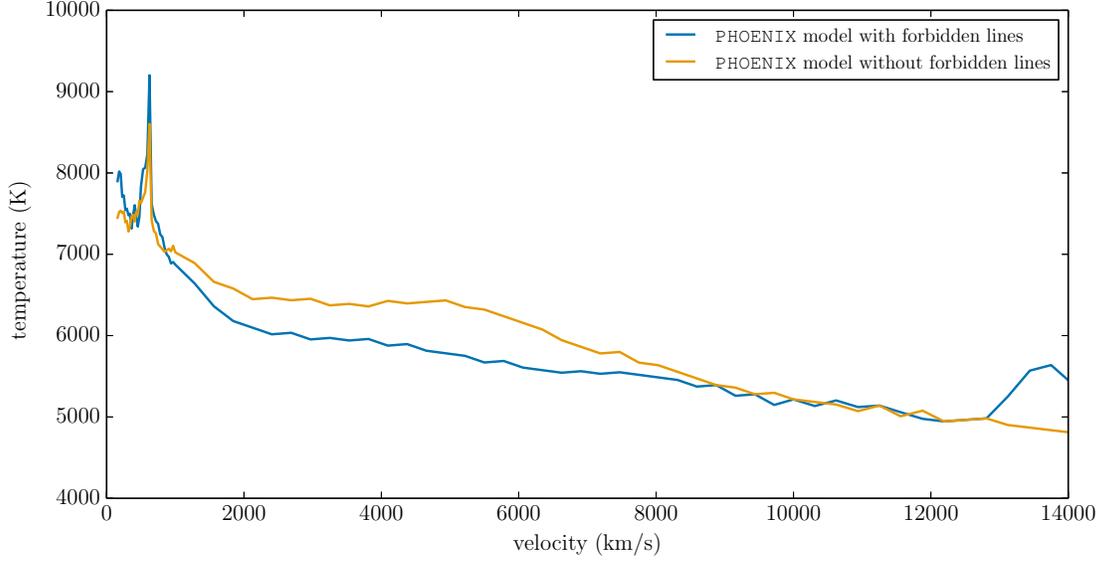}
    \caption{Converged temperature structures for the models with (blue) and without (orange) forbidden lines.
        These lines act as coolants and tend to lower the temperature \citep{dessart14}.}
    \label{fig:temps}
\end{figure}

\begin{figure}
    \epsscale{1.0}
    \plotone{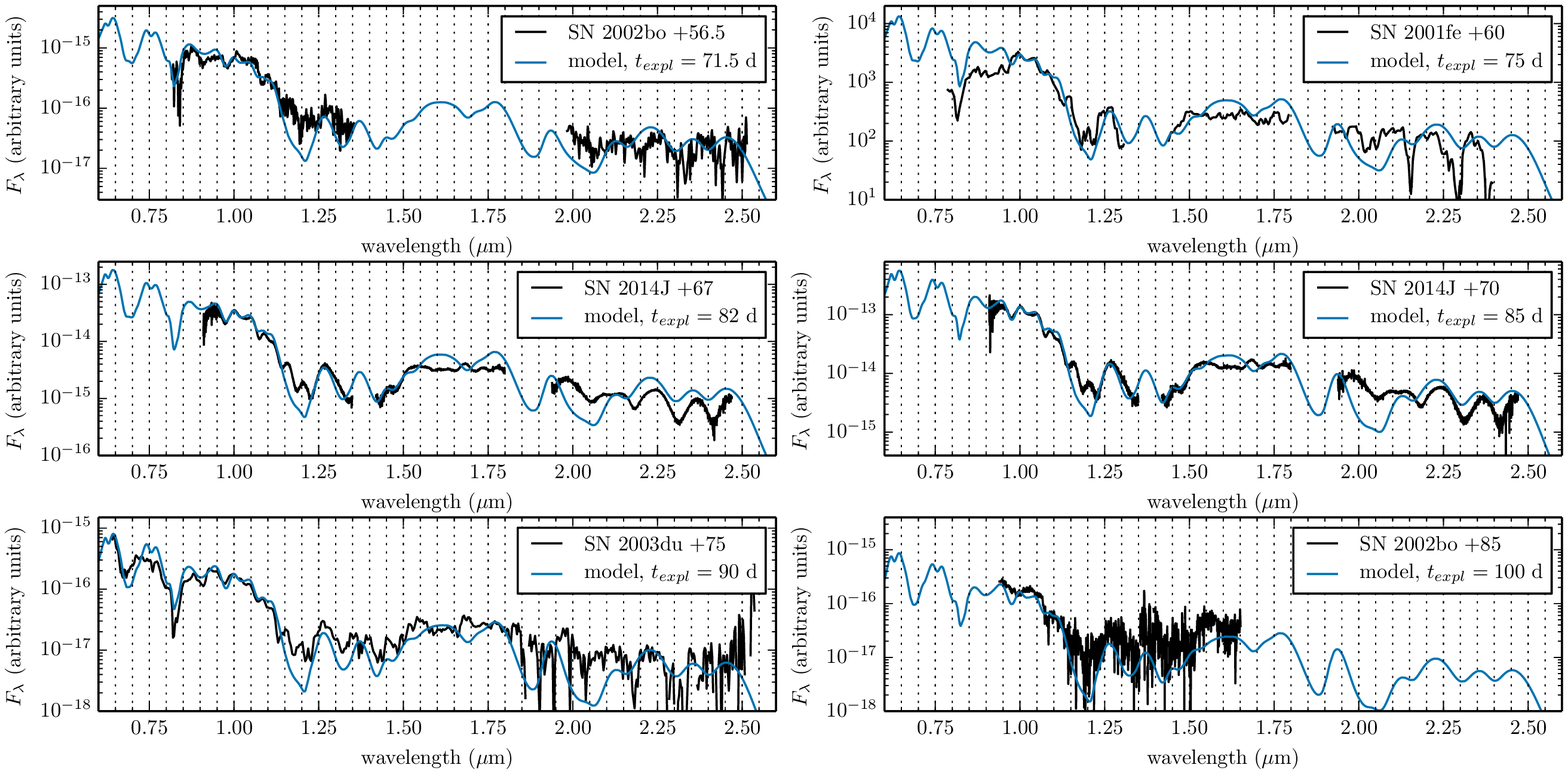}
    \caption{\texttt{PHOENIX} spectra with forbidden lines compared to all observed SN spectra in this work.
      The homologous expansion time for each synthetic spectrum was 15 days, such that, e.g., the model spectrum corresponding to SN~2001fe +60 had an expansion time of 75~d.}
    \label{fig:phx_forb_vs_all_obs}
\end{figure}

No ion in the permitted-line-only spectrum in Figure~\ref{fig:phx_single_ion_no_forb_vs_14J} appears able to reproduce the emission at 1.98~$\mu$m.
However, addition of forbidden line data produces a strong emission feature in the synthetic spectrum near that wavelength.
Figure~\ref{fig:phx_single_ion_forb_vs_14J} indicates that the emission is due to [\ion{Ni}{2}], which we identify as $\lambda1.939~\mu\text{m}$.
This emission feature produced in the synthetic spectrum is too blue by about 0.5~$\mu$m compared to the corresponding feature in SNe~2014J (Figure~\ref{fig:phx_single_ion_forb_vs_14J}).
Curiously, this shortcoming is repeated in all of the other SNe in our sample (see Figure~\ref{fig:phx_forb_vs_all_obs}).
The spectral coverage of SN~2002bo at both day +56.5 and day +85 unfortunately does not span this region and so we cannot study the influence of [\ion{Ni}{2}] in that object.
An explosion model which leaves the $^{58}$Ni more concentrated at the center may cause the nickel emission to shift redward slightly, providing a better match to the three SNe~Ia with available spectral coverage studied in this work.

This [\ion{Ni}{2}] emission appears in our SN Ia sample to be consistently redward of its appearance in our models.
In addition, its location in our model spectra does not change over the 30-day time frame spanned in this work: the emission center is at $1.94$~$\mu$m, very near the center of the line rest wavelength.
It is possible that the systematic redshift of this feature in our sample of observations is due to to asymmetrical explosions combined with a line-of-sight effect, as Doppler shifts of optically thin emission lines are sensitive to hydrodynamical asymmetries \citep{maeda10}.
If asymmetry is indeed the explanation for this observed redshift, then this feature may possess great utility as a probe of asymmetry at quite early epochs of a SN~Ia, assuming a complete and high-quality series of NIR spectra can be obtained.

Given the model age of $t_{\text{expl}} = 82$~d, the radioactive $^{56}$Ni, with a half-life of $\sim 7$~d, has all but disappeared.
Therefore the [\ion{Ni}{2}] which produces the emission feature at this epoch (the feature is present in varying degrees in all SNe in Figure~\ref{fig:01fe_03du_14J} except SN~2002bo, which lacks the necessary spectral coverage) must instead be a stable isotope, specifically $^{58}$Ni, which is a product of high-density nuclear burning \citep{thielemann86,iwamoto99}.
The [\ion{Ni}{2}] emission in the series of model spectra shown in Figure~\ref{fig:phx_forb_vs_all_obs} exhibits similar behavior to, e.g., the [\ion{Fe}{2}] and [\ion{Fe}{3}] lines which appear in optical spectra in the nebular phase \citep[e.g.,][]{leloudas09}, strengthening as the optical depth of the inner regions of the ejecta decreases and more of the central $^{58}$Ni is exposed.

Although the wavelength of the [\ion{Ni}{2}] emission in Figure~\ref{fig:phx_forb_vs_all_obs} does not change over the 30 days spanned by the models, it does strengthen noticeably.
We estimate then that a series of moderate-cadence (periods of $\sim 20$ days) NIR observations of individual SNe~Ia during these epochs would provide the information necessary to confirm or refute this identification of $^{58}$Ni in this particular hydrodynamical model.
(Observations with longer periods would also resolve the strengthening of this emission, but may be unable to determine the time scale over which it evolves.)
They would likely also allow models to constrain the amount of $^{58}$Ni produced in the explosion, since the strength of the [\ion{Ni}{2}] emission should vary with the mass of $^{58}$Ni.
(In this particular model the mass ratio of stable $^{58}$Ni to radioactive $^{56}$Ni is $\sim 0.15 : 1$, a typical value for delayed-detonation explosion models.)
If the $^{58}$Ni identification is correct, observations with such a cadence would be able to track the Doppler shift of this feature (or lack thereof), thereby constraining its velocity extent.
However, given this emission feature's close proximity to a heavily polluted telluric region, the sites for such observations would need to be selected carefully, i.e., at high elevation.
SNe~Ia at higher redshift would exhibit a lower signal-to-noise ratio, but this emission feature in those objects would be less susceptible to telluric contamination.

Besides $^{58}$Ni, Mn is also a product of high-density burning, and so if the (tentative) identification of \ion{Mn}{2} at 1.15~$\mu$m is correct, this would provide a second indication that such burning took place in SNe 2001fe, 2002bo, 2003du, and 2014J, presumably during the deflagration phase of the delayed-detonation explosion.
\citet{marion14} showed that the layered structure apparent in the early optical and NIR spectra of SN~2014J is consistent with a delayed-detonation explosion model.
In addition, \citet{churazov14} found that the mass and distribution of radioactive $^{56}$Ni in SN~2014J is consistent with a near-Chandrasekhar-mass explosion, consistent with our results.
Sub-Chandrasekhar explosion models tend to underproduce neutron-rich elements such as $^{58}$Ni \citep{seitenzahl13}, and so the NIR \ion{Mn}{2} and [\ion{Ni}{2}] spectral signatures presented here may provide useful proxies for diagnosing near-Chandrasekhar-mass explosions.

\section{Conclusions}
\label{sec:conclusions}

We have presented NIR synthetic spectra of a SN~Ia delayed-detonation hydrodynamical model at epochs spanning from $t_{\text{expl}} = 71$~d through $t_{\text{expl}} = 100$~d, corresponding roughly to the ages of the NIR spectra of SNe~2001fe, 2002bo, 2003du, and 2014J spanning from days +56.5 through +85, which at these epochs are undergoing the transition from the photospheric to nebular phase.
We found that our model reproduces most features in the NIR spectrum with permitted lines of \ion{Fe}{2} and \ion{Co}{2}, confirming the results of \citet{branch08,gall12}.
Only a few other species contribute appreciably to the NIR spectrum, including \ion{Ni}{2}, \ion{Co}{2}, \ion{Ca}{2}, and \ion{Mn}{2} at 1.15~$\mu$m, and \ion{Ca}{2} and possibly \ion{Ni}{2} at 1.20~$\mu$m.
The emission at 1.98~$\mu$m seen in three of the four SNe~Ia studied here is the only feature which arises from forbidden lines in our model; it is due to forbidden lines of [\ion{Ni}{2}].
That the emission is so strong indicates a significant mass of stable $^{58}$Ni in the ejecta (our model has 0.09~$M_\odot$), produced during the early stages of the explosion of the WD progenitor.
The evidence for [\ion{Ni}{2}] and possibly of \ion{Mn}{2} lines implies burning at high density, suggesting that the progenitors of these three SNe~Ia ignited at nearly the Chandrasekhar mass.

\acknowledgments
We are grateful to the referee for a thoughtful review which improved significantly the quality of this work.
This research has made use of NASA's Astrophysics Data System and the Weizmann Interactive Supernova Data Repository (WISeREP), and is based on observations obtained with the Apache Point Observatory 3.5-meter telescope, which is owned and operated by the Astrophysical Research Consortium.
This work has been supported in part by support from NSF grant AST-0707704 and by support for programs HST-GO-12298.05-A, and HST-GO-122948.04-A provided by NASA through a grant from the Space Telescope Science Institute, which is operated by the Association of Universities for Research in Astronomy, Incorporated, under NASA contract NAS5-26555.
This research used resources of the National Energy Research Scientific Computing Center (NERSC), which is supported by the Office of Science of the U.S. Department of Energy under Contract No.  DE-AC02-05CH11231; and the H\"ochstleistungs Rechenzentrum Nord (HLRN).
We thank both these institutions for a generous allocation of computer time.

{\it Facilities:} \facility{ARC}

\begin{deluxetable}{lcccc}
    \tablecaption{Summary of Objects\label{objsum}}
    \tablehead{\colhead{Name} & \colhead{$M^{\text{max}}_B$} & \colhead{$\Delta m_{15}$} & \colhead{$v_{\text{Si~\textsc{II},max}}$} & \colhead{Spectroscopic} \\ & & & (km~s$^{-1}$) & Classification}
\startdata
SN~2001fe & ---    & ---  & $11,200$ & NV \\
SN~2002bo & $-19.41$ & $1.13$ & $12,800$ & HV \\
SN~2003du & $-19.34$ & $1.04$ & $10,500$ & NV \\
SN~2014J  & $-19.19$ & $1.11$ & $11,900$ & HV \\
\enddata
\vspace{-0.3in}
\tablecomments{Summary of photometric and spectroscopic characteristics of the four SNe~Ia studied in this work.
    $M^{\text{max}}_B$ is the absolute magnitude in the $B$ band at maximum light in $B$.
    $\Delta m_{15}$ is the number of magnitudes by which the object declines in $B$ between maximum light and 15 days layer \citep{phillips93}.
    (Photometry of SN~2001fe is unavailable.)
    $v_{\text{Si~\textsc{II},max}}$ is the velocity of the \ion{Si}{2}~$\lambda6355$ absorption feature at maximum light in $B$.
    The spectroscopic classification consists of ``normal velocity'' (``NV'') and ``high velocity'' (``HV''), based on $v_{\text{Si~\textsc{II},max}}$, as defined in \citet{wang09}; the division between the two regimes is $\sim 11,800$~km~s$^{-1}$, placing SN~2014J barely in the HV group.}
\end{deluxetable}

\begin{deluxetable}{lccccc}
\rotate
\tabletypesize{\small}
\tablecaption{Summary of Observations\label{obssum}}
\tablehead{\colhead{UT Date} & \colhead{Object} & \colhead{Facility} & \colhead{Exposure Time} & \colhead{Total On-Source Integration Time} & \colhead{Airmass} \\ & & & (sec) & (sec) & }
\startdata
MJD 56755.14 & SN 2014J & APO 3.5m/TripleSpec & 300 & 3000 & 1.25 \\
MJD 56755.17 & HD 63586 & APO 3.5m/TripleSpec & 7 & 112 & 1.18 \\
MJD 56758.33 & SN 2014J & APO 3.5m/TripleSpec & 300 & 1500 & 1.59 \\
MJD 56758.36 & 39 UMa & APO 3.5m/TripleSpec & 20 & 240 & 1.51 \\
\enddata
\vspace{-0.3in}
\tablecomments{The mid-exposure UT dates (in Modified Julian Date) for our observations of SN 2014J and corresponding A0V-star calibrators are tabulated, along with the duration of each individual exposure at an ABBA nod position and total on-source integration time are compiled.}
\end{deluxetable}

\end{document}